\documentclass{article}

\usepackage[table]{xcolor}
\usepackage{microtype}
\usepackage{graphicx}
\usepackage{subfigure}
\usepackage{booktabs} 

\PassOptionsToPackage{hyphens}{url}\usepackage{hyperref}

\usepackage[accepted]{icml2024}

\usepackage{amsmath}
\usepackage{amssymb}
\usepackage{mathtools}
\usepackage{amsthm}
\usepackage{nicematrix}
\usepackage[capitalize,noabbrev]{cleveref}

\theoremstyle{plain}

\theoremstyle{definition}

\theoremstyle{remark}

\usepackage[textsize=tiny]{todonotes}

\usepackage{tabularx}

\icmltitlerunning{AI-Powered Autonomous Weapons Risk Geopolitical Instability and Threaten AI Research}

\begin{document}

\twocolumn[
\icmltitle{
AI-Powered Autonomous Weapons Risk Geopolitical Instability and Threaten AI Research}



\icmlsetsymbol{equal}{*}

\begin{icmlauthorlist}
\icmlauthor{Riley Simmons-Edler}{equal,harvard,kempner}
\icmlauthor{Ryan P. Badman}{equal,harvard,kempner}
\icmlauthor{Shayne Longpre}{mit}
\icmlauthor{Kanaka Rajan}{harvard,kempner}

\end{icmlauthorlist}

\icmlaffiliation{harvard}{Department of Neurobiology, Harvard Medical School, Boston, USA}
\icmlaffiliation{mit}{Massachusetts Institute of Technology, Cambridge, USA}
\icmlaffiliation{kempner}{Kempner Institute, Harvard University, Cambridge, USA}

\icmlcorrespondingauthor{Riley Simmons-Edler}{riley\_simmons-edler@hms.harvard.edu}

\icmlkeywords{Machine Learning, ICML, AI Policy, AI and Society, Military AI, AI Safety}

\vskip 0.3in
]



\printAffiliationsAndNotice{\icmlEqualContribution}

\begin{abstract}
The recent embrace of machine learning (ML) in the development of autonomous weapons systems (AWS) creates serious risks to geopolitical stability and the free exchange of ideas in AI research.
This topic has received comparatively little attention of late compared to risks stemming from superintelligent artificial general intelligence (AGI), but requires fewer assumptions about the course of technological development and is thus a nearer-future issue.
ML is already enabling the substitution of AWS for human soldiers in many battlefield roles, reducing the upfront human cost, and thus political cost, of waging offensive war.
In the case of peer adversaries, this increases the likelihood of ``low intensity'' conflicts which risk escalation to broader warfare.
In the case of non-peer adversaries, it reduces the domestic blowback to wars of aggression.
This effect can occur regardless of other ethical issues around the use of military AI such as the risk of civilian casualties,
 and does not require any superhuman AI capabilities.
Further, the military value of AWS raises the specter of an AI-powered arms race and the misguided imposition of national security restrictions on AI research.
Our goal in this paper is to raise awareness among the public and ML researchers on the near-future risks posed by full or near-full autonomy in military technology, and we provide regulatory suggestions to mitigate these risks.
We call upon AI policy experts and the defense AI community in particular to embrace transparency and caution in their development and deployment of AWS to avoid the negative effects on global stability and AI research that we highlight here.

\end{abstract}

\section{Introduction}
\label{sec:intro}

With the rise of powerful generative AI models such as GPT-4 and Stable Diffusion, and continued progress in fields such as robotics and reinforcement learning, concerns have grown among both experts and the public about giving AI too much power.
Academic concerns have thus far centered on threats in domains such as cybersecurity, biological weapons, disinformation, fraud, and hypothetical rogue artificial general intelligence (AGI) \cite{Future_of_Life_Institute2023-lj,Bengio2023-wn}.
Despite these general concerns, there has been relatively little attention given to specific recent developments from military and defense-industry groups, which have already begun to deploy next-generation AI-guided Autonomous Weapon Systems (AWS).
Weapons falling under the AWS label have traditionally been either remotely operated (but AI-assisted) or only autonomous within a very narrow scope.
However, new fully autonomous (unmanned) AWS in development remove or reduce the human role in the control and decision making process, with the goal of removing humans from the active battlefield en mass. 
For example, the Pentagon's Replicator program for AI-based weapon ``swarms'' promises a drastic shift in warfare towards highly autonomous and cooperative AI units within the next few years \cite{Greenwalt2023-io}. 
AWS can serve many battlefield roles, although human-targeting lethal AWS (LAWS) have received most public attention \cite{Taddeo2022-hu}.
Many of these new advanced AI and machine learning (ML) weapons systems are already seeing real-world deployment for the first time in the Ukraine War, and are being designed for every branch of the military and by many nations \cite{Sharma2022-mb, Greenwalt2023-io, Anthony_Pfaff2023-sr}. 

Current discussion of ethical use of AWS centers on keeping humans in-the-loop for remote autonomous warfare so that only humans are making lethal decisions, while accepting the idea that remote warfare saves lives \cite{Riesen2022-fj, Anthony_Pfaff2023-sr}. 
However, we argue that there are fundamental issues caused by removing humans from the battlefield.
Human ``boots-on-the-ground'' can signify a commitment to following the rules of war, improve humanitarian aspects of occupation, and most importantly maintain a human cost to war for aggressor nations that prevents a state of endless war from being politically feasible \cite{Trapp2018-dw}. 
We are concerned the recent embrace of AWS by global militaries is leading to a future where wars are more frequent, with such warfare having negative consequences for global stability even if AWS reduce civilian casualties relative to human soldiers.
This new model of AWS-centered warfare will be supported by an increasing fusion of civilian and military AI research that will have devastating effects on research and trust in our field.
Thus, we have written this article to make ML researchers aware of the rapid pace of AWS development and the need for action, and to provide community and policy suggestions to help prevent this possible future. 

The ethical implications of such AWS capable of selecting and destroying targets without meaningful human control are considerable, and have been discussed by prior work at length \cite{Scharre2018-oq, Blanchard2022-un, Brown2023-wf}.
Over the past 2-3 decades many voices have called for these systems to be banned entirely and/or governed by international treaties \cite{Russell2023-qx, Bode2023-yu}.
We agree that the ideal would be that such systems were not developed, due to moral hazards in their use as well as the issues we highlight here.
However, in this paper we focus on the scenario in which a middle ground is achieved: AWS are not banned despite moral reasons they ought to be \cite{Geist2016-kn}, but manage to comply with basic military ethics and do not increase civilian casualties compared to conventional war \cite{Scharre2018-oq, Brose2020-ov, Riesen2022-fj}.
This assumption that AWS will not increase civilian casualties is an optimistic one ---likely many AWS will fall short of that goal---
but it provides the most conservative starting point for our discussion on the indirect negative impacts of AWS.
In 2024, major military powers have already committed to making autonomous weapon systems core components of their armed forces, and thus the time to discuss full AWS bans seems increasingly in the past \cite{Greenwalt2023-io, Klare2023-pm, Brose2020-ov}. 
The conversation now needs to center on responsible development and proliferation of the rapidly improving AWS systems that are already being developed and used by global militaries.

Official statements and announced R\&D projects make it clear that the direction of AWS development efforts both short- and long-term is the removal of human soldiers from direct combat roles, to reduce casualties and increase combat effectiveness \cite{Greenwalt2023-io, Anthony_Pfaff2023-sr, Bachmann2023-tu}.
While these goals are reasonable in isolation, a lack of public attention and transparency around the rapid and increasing pace of AWS development and employment risks humanity sleepwalking into an AWS arms race between global powers.
We will see these risks in the near future. China and Russia have given 2028-2030 as targets for major automatization of their militaries to begin, and while the USA is set to begin deployment sooner, the country is planning for a long-term transformation \cite{Sharma2022-mb, Greenwalt2023-io, Kania2020-dy, Warren2022-hb}.

\begin{table*}[t!]
  
    \centering
    \begin{tabularx}{\textwidth}{|>{\hsize=.12\hsize}X|>{\hsize=.44\hsize}X|>{\hsize=.44\hsize}X|}
        \hline
         \textbf{\textsc{Topic}} & \textbf{\textsc{Challenges}} & \textbf{\textsc{Recommendations}} \\ \hline \hline
         Major power aggression & AWS replacement of human soldiers makes war more domestically palatable.
         Defenders turn to asymmetric warfare/terrorism for deterrence.
         & Require significant human battlefield presence, focus on human-AWS teaming over remote and fully autonomous AWS-centered conflict.\\ \hline \rowcolor[gray]{0.9}
         Peer conflict escalation & Heavy use of AWS makes conflict initiation easier between major powers, risks AWS arms race. & International transparency about broad capabilities, deployment disclosures for AWS systems. \\ \hline
         Conflict transparency & Lack of human presence means accountability in war is harder, war crimes and battlefield under-performance less visible to leaders and the public. & Require detailed public reports on AWS capabilities, deployments, and outcomes.
         Embed oversight/watchdogs in AWS command centers.\\ \hline \rowcolor[gray]{0.9}
         Proliferation & Development and sale of AWS will be widespread, global availability of AWS is inevitable. & Avoid futile AI hardware/software restrictions.\\ \hline
         AI researcher restrictions & Military AI needs lead to censorship of civilian research, reduced international collaboration, monitoring and restriction of researchers. & Universities, corporations, governments, etc. establish norms on how much military and civilian research should overlap. 
         \\ \hline \rowcolor[gray]{0.9}
         Dual-use AI technology & Many AI algorithms are innately dual-use.
         Facial recognition, navigation, robotics, etc.
         Military interest in civilian research is likely to grow. & Improve university ethics oversight and transparency for military-funded AI research, and caution researchers against efforts to weaponize AI. \\ \hline
         Risk of over-regulating AI & Public backlash to AWS leads to calls for more limitations on AI research generally, hurting international research community and academic norms. & Avoid restricting basic AI research, regulate explicit AWS research and military-related datasets over general civilian hardware and AI models.\\ \hline \rowcolor[gray]{0.9}
         Autonomy transparency & Public data on current AWS are often vague on autonomy levels, definitions of human-in-the-loop, may be more autonomous in practice. & Require governments and AWS manufacturers to clarify the degree of autonomy of AWS.
         Set international standards for allowed levels of autonomy.\\
         \hline
    \end{tabularx}
    \caption{Overview of AWS issues raised and policy recommendations in this work.}
    \label{tab:summary}
   
\end{table*}

Given these priors, we argue that because highly capable AWS lower the human costs associated with conflict initiation and escalation, they also create a large risk to global geopolitical stability. 
Critically, this effect worsens the more capable AWS become, even if civilian casualties and collateral damage decrease, and cannot be solved by simply improving the ML systems involved-- deliberate policy actions are needed.

In conflicts between powers with large disparities in military strength, invasion or intervention using an AWS-heavy force is politically easier than with an all-human force, since there will be fewer deaths on the aggressor side \cite{Moreau2011-qx, Greenwalt2023-io}.
However, reducing human casualties in a single conflict can be outweighed if the number of conflicts that occur increases. 
Indeed, the past century suggests that when dominance in a new military technology leads to regional or global hegemony, this does not always translate to greater stability, and can actually increase lower intensity conflicts and terrorism \cite{Benvenisti2004-ji, Drezner2013-de, Hegre2014-ye}.
AWS-heavy armies with minimal human battlefield presence may lead to a rise in terrorism, assassination, attacks on civilians, and other methods extending beyond the traditional battlefield \cite{Kwik2022-fh}.
These abhorrent methods provide a way for less powerful nations who lack AWS to deter or retaliate against nations deploying AWS-heavy forces despite their inability to do so through battlefield casualties \cite{Moreau2011-qx}. 

In addition, AWS that function as a substitute for ``boots on the ground'' are a threat to internal stability and representative government, unless careful safeguards are in place to protect against tyrannical use or sabotage of such systems \cite{Ams2023-hx}.
Further, reduced human battlefield presence makes journalistic transparency and civilian oversight of conflict more difficult, and makes it easier to hide war crimes and the impacts of war from both civilian leadership and the public \cite{Lin-Greenberg2020-lf, Ams2023-hx}.

Beyond the impacts AWS have on global stability, the importance of AWS for warfare will likely lead to major negative impacts on civilian AI research. 
AWS will become a revolutionary military technology, as did nuclear weapons, mechanized warfare, and others historically, but with important differences in the ease of AWS proliferation and impact on civilian technology development \cite{Stern2023-ql, Johnson2020-rt, Laird2020-ig, Wong2020-oq}.
Recent work by Schneider argues that the prevailing military logic\textemdash that hegemonic power creates stability\textemdash can be highly erroneous when that power is based on technology that relies on a scarce or controlled resource, the capability/vulnerability paradox \cite{Schneider2019-nj}.
In the case of AWS, these key resources are AI experts and knowledge, access to relevant big data, and semiconductor manufacturing\textemdash all of which have historically been dominated by a handful of countries \cite{Chu_undated-gx, Daniels2019-ib, Saif_M_Khan_Alexander_Mann_Dahlia_Peterson_undated-pn, Farrell2022-go, Flynn2020-rk}.
While these resources are becoming more available globally, governments with a current advantage are likely to try to secure that advantage by preventing their spread.
We may see a rise in export restrictions, publication oversight and redaction, and knowledge compartmentalization in the field of AI as a result \cite{Bengio2023-wn, Rickli2021-nj, Warren2020-ey}. 

Longer term, we could see AI resources treated analogously to how oil resources have fueled conflict since the advent of industrialized warfare \cite{Schneider2019-nj}.
Indeed, we see hints in current events that this misguided idea is spreading \cite{Mastro2023-hy, Klingler-Vidra2023-di, Triolo2023-cy}, and some militaries are already seeking to use civilian AI experts to speed development of domestic AWS \cite{Anthony_Pfaff2023-sr}. 
China for example has declared Military-Civil Fusion as foundational to updating its military with AWS capabilities, and major Chinese universities have been roped into AI weapons programs \cite{Margarita_Konaev_Ryan_Fedasiuk_Jack_Corrigan_Ellen_Lu_Alex_Stephenson_Helen_Toner_Rebecca_Gelles2023-qv, Can2022-pe, Warren2021-nz, Fedasiuk2021-fj}.

To reduce and prevent these outcomes, action by AI researchers, policymakers, and the public will be needed.
Below we propose several policies and actions to take in Section \ref{sec:policy} and Table \ref{tab:summary}, and briefly review highlights here.
First, the best way to avoid AWS-driven conflicts is for nations to avoid developing or fielding AWS that operate independently rather than in cooperation with humans, especially in large scale or prolonged conflicts.
Second, the international community needs to build consensus around AWS autonomy levels, and which levels are acceptable or unacceptable and in what circumstances.
Third, there is an urgent need for transparency around AWS capabilities and levels of human oversight in actual use\textemdash
``Human-in-the-loop'' is becoming a meaningless phrase that arms manufacturers and militaries use to reassure the public.
Today, ``human-in-the-loop'' is shifting from implying hands-on remote control to automated targeting where a human presses a ``go'' button with minimal insight into what the AWS is doing or why \cite{Bode2023-yf, Bode2023-yu, Boys2023-aa, Abraham2024-wy}.
Fourth, universities and companies worldwide must draw lines and declare publicly what level of military-civilian overlap and collaboration on AWS-related research is acceptable \cite{DAgostino2024-pw}.



The structure of this article is as follows: Section \ref{sec:current_aws} discusses the state of AWS development and deployment today, as well as near-future directions.
Section \ref{sec:proliferation} discusses why limiting AWS proliferation is difficult, and the ineffectiveness of restricting access to ML hardware and expertise to prevent it.
Section \ref{sec:policy} discusses several approaches to mitigate the risk posed by advanced AWS short of a total ban.

\section{Current State of AWS}
\label{sec:current_aws}

The past century has seen the development of mechanized, remote-controlled, and semi-autonomous weapons of war as extensively covered in other work \cite{Scharre2018-oq, Work2021-cs}.
However, the past 5-10 years have seen unprecedented progress in and acceptance of truly autonomous weapons systems that are increasingly able to hunt and attack targets with minimal or no human oversight.
We review the current state of these truly autonomous (i.e. unmanned, uncrewed) weapons systems for land, air and sea in this section to provide context on the likely trajectory AWS development will take. Note that we omit discussion of outer space-based AWS as there is currently too little information publicly available \cite{Van_Esch2017-uk}.
In addition, to our knowledge no work has comprehensively reviewed AWS types across combat domains in an up-to-date way.
Current major reviews have focused on AWS progress for only one branch of the military, and/or do not include the most recent developments in this fast-moving field \cite{Sharma2022-mb, Castillo2022-tk, Longpre2022-ht, Bode2023-yf}.

Note that we do not discuss the impacts of AI on cyber warfare in this work, as it is a complex topic worthy of further study that has distinct impacts compared to directly lethal AWS.
Nonetheless it is a rapidly developing field of warfare, with both China and Russia proclaiming it a major topic of military development \cite{Fedasiuk2021-fj, Thornton2020-qa,Dewey2021-rp,Takagi2022-hv}.

\subsection{Airborne AWS}
\label{subsec:air_aws}

One of the most established and battle-tested AWS types is loitering munitions (LMs) \cite{Bode2023-yf}.
Loitering munitions include a wide range of designs, but are single-use unmanned aerial combat units which can hover or circle, perform detection and targeting, and then autonomously launch (crash) themselves into targets with lethal payloads. 
Typically LMs do not require human control or oversight after launch/takeoff, and prosecute their pre-programmed missions autonomously.
The main operational LM unit up until approximately 2015 was the Israeli Harpy \cite{Scharre2018-oq, Scharre2015-kx}, which can autonomously detect and destroy radar targets in an area whose precise locations were not known before launch, and can return to base and be re-used if no such targets are found.

Since then, the number of LMs has rapidly expanded, with at least 24 models of varied size and sophistication across 16 countries, and the capabilities of the AI systems used by LMs have increased considerably \cite{Bode2023-yf}.
Most concerning, the range of targets for LMs has expanded from radar systems to armored vehicles and even specific human personnel that can be identified and targeted by facial recognition without operator involvement.
The Kargu-2 developed by Turkey is one of the foremost systems with this capability, and was deployed on battlefields in Libya \cite{Longpre2022-ht, Bode2023-yf}.
Other notable models such as the Rafael Spike Firefly and the IAI Rotem L are optimized for targeting in urban environments, and at least 14 other models have specifications suggesting they can autonomously target and follow or execute individual human targets \cite{Bode2023-yf}.
Both Ukraine and Russia have deployed LMs of various levels of sophistication and autonomy in their current war in large numbers \cite{Bode2023-yf, Cotovio2024-vt}.
Notably, the former commander in chief of the Ukrainian military recently highlighted LMs and anti-LM defenses such as jamming as critical to their war effort \cite{Zaluzhnyi2023}. 
He also highlighted the Russian Lancet LM, which uses a western-built Nvidia Jetson TX2 ML compute module to autonomously track targets, as being particularly difficult to defend against \cite{Zaluzhnyi2023,Faragasso2023-uy}.
A major issue in assessing LMs is that governments and manufacturers are often vague about the capabilities of LM AI-targeting systems, and whether autonomous modes without human oversight have been used operationally \cite{Bode2023-yf, Hambling2023-ps, Sayler2021-xa, Longpre2022-ht}.

The other well-established form of AWS is the (reusable) unmanned aerial vehicle (UAV).
A major recent announcement in AWS-UAV development is the Pentagon's Replicator Program, which aims to deploy thousands or tens of thousands of UAVs within 18-24 months, across land, air, and sea \cite{Klare2023-pm, Greenwalt2023-io}.
The prime candidates for these programs include budget fighter and bomber drones that can replace traditional crewed aircraft but with a 10-100x lower price tag. 

One example system is the XQ-58A Valkyrie developed by Kratos Defense and Security Solutions to be a ``loyal wingman'' to manned aircraft, defending them and performing offensive actions that would otherwise risk the manned aircraft, first demonstrated in 2019 \cite{Wang2020-vs, Klare2020-wc, Klare2023-pm, Newdick2023-hb}.
Notably, Kratos has claimed it can produce several hundred Valkyrie units per year at per-unit costs between \$2-5 million, cheaper than manned aircraft, most traditional drones like the American Reaper UAV, and even some missiles \cite{Jordan2021-tv, Lyu2022-sp, Boys2023-aa}.
These units can scout, are equipped with weapons for defensive fire, have an operational range of ~3000 miles, and can even deploy smaller unmanned aircraft for strike or low-level reconnaissance \cite{Newdick2023-hb}.
Other major candidates are even cheaper, and are intended to be deployed in large numbers using autonomous swarming capabilities, such as Shield AI’s V-Bat and Boeing’s MQ-28 \cite{Gruszczak2023-eq, Saylam2023-zm, Bachmann2023-tu, Sud2020-ss, Castillo2022-tk}.
In addition, unmanned versions of the F-16 have been demonstrated recently, 
with deployment estimated for 2025 \cite{noauthor_undated-fx, Brown2023-wf, Kohler2024-oa}. 
US allies are developing similar systems, such as the Blue Bear Ghost line of autonomous UAVs \cite{noauthor_2018-mo}. 

China is also making considerable progress in improving the autonomous capabilities of UAVs in the People’s Liberation Army-Air Force (PLAAF).
The Wing Loong line of unmanned aerial vehicles (UAVs) was declared to be a focus for autonomous upgrades to minimize the need for a remote operator, using 5G, AI, and big data \cite{Liu_Xuanzun_undated-hf}.
Looking forward, China recently announced that a significant degree of autonomy (usually translated as ``intelligence'') must be introduced to the military over the coming decades \cite{Kania2020-dy, Warren2022-hb}.

\subsection{Ground AWS}
\label{subsec:ground_aws}

The USA and other NATO nations have driven most recent UGV developments of note \cite{Sharma2022-mb}.
In 2017, the US Army declared its mid-term (2021–30) priorities include achieving the ability to perform fully automated convoy operations using unmanned combat and support vehicles that can navigate difficult terrain and carry advanced payloads \cite{US_Army_Training_and_Doctrine_Command2017-uf}.
The most notable recent unmanned ground vehicle (UGV), already deployed on the Ukrainian battlefield \cite{Robotics2023-bd}, is the Milrem THeMIS unit, which is capable of autonomous targeting and ordinance deployment as well as other support and sensing tasks \cite{Mathiassen2021-lg, noauthor_2017-lp}.
Milrem has sought to make the unit more autonomous for faster threat response and to enable coordination in autonomous swarms.

Several other UGV AWS are in late development stages across the US, Europe and India, with major demonstrations or deployments planned \cite{Sharma2022-mb}.
For example, the MAARS Grey Muzzle autonomous weaponized robot developed by QinetiQ that is equipped with machine guns and grenade launchers for direct fire support roles.
This system features an onboard AI system for autonomous navigation and waypoint following, but the weapons require human input before firing \cite{noauthor_2023-ww}.
Legged quadruped UGVs also have shown great promise for use in more complex terrain and environments \cite{noauthor_undated-bm, Gilead2023-wq}.
Lastly, Pegasus Intelligence is developing a portfolio of diverse AWS, some of which are amphibious or can transition between flight and ground modes \cite{Pegasus-Intelligence2023-wk}. 
Looking forward, recent progress in subfields of ML and robotics, such as autonomous vehicle research, has greatly improved UGV capabilities and spurred additional development, with several recent DARPA programs testing self-driving abilities in military technology \cite{Bisht2022-qa}.

\subsection{Naval AWS}
\label{subsec:naval_aws}

Submarine AWS development has been a high priority for navies, as the ocean is an inherent signal jammer, making remote operation difficult or impossible\textemdash indeed, modern naval mines and torpedoes can be seen as very simple AWS.
One of the most recent notable developments in naval AWS was the 2023 Israel Aerospace Industries (IAI) announcement that its BlueWhale anti-submarine warfare (ASW) uncrewed submarine has participated in NATO exercises \cite{noauthor_2023-ti}.
BlueWhale ASW can operate for 2-4 weeks with satellite connection and electro-optical sensors for sea targeting, with low operational cost.

Comparable efforts are being made by the US Navy.
An array of medium, large and extra-large sized unmanned naval vehicles are planned for undersea and surface operations \cite{ORourke2023-fv}.
The largest of these is the Orca XLUUV, with a claimed dry weight in excess of 50 tons. It specializes in Hammerhead mine deployment on the seabed but is reconfigurable depending on needs for combat, surveillance, intelligence, etc. and is capable of carrying multiple types of payloads \cite{ORourke2023-fv}.
L3Harris is also developing a line of autonomous surface vessels up to 13 meters long which require no crew for operation and only minimal remote operation for payload delivery \cite{noauthor_2023-om, noauthor_undated-fd}.
These new surface vessels are likely much more capable and autonomous than the smaller semi-autonomous surface vessels which have seen notable use in the Ukraine War \cite{Ozberk_undated-bu, Santora2023-wf}.
For detailed historical progress on autonomous units in the US Navy, see \cite{Castillo2022-tk}. 

China has greatly advanced its claimed unmanned surface and undersea capabilities the past 5 years. The country has announced plans for the world's first AI-run deep sea underwater colony to house and oversee submarine operations for defense and survey in the South China Sea \cite{Chen2018-to}.
By 2021, an estimated 48 universities and 45 enterprises in China were involved in unmanned (UUV) and autonomous undersea vehicle (AUV) R\&D projects \cite{Fedasiuk2021-fj}.
Prototype units include large and small submarines and surface vessels, and military project descriptions mention deep learning-based targeting and sensor analysis, as well as mine deployment \cite{Fedasiuk2021-fj, Goldstein_undated-po}.
Disentangling civilian versus military models is increasingly difficult, as the Tianjin Municipal Government for example promotes dozens of AUV and UUV models as being capable of both civilian and military use \cite{Fedasiuk2021-fj}.

\subsection{AWS Systems for Command}
\label{subsec:aws_command}

Data and intelligence sharing and synthesis are obvious targets for AI enhancement and AWS integration. A number of efforts to build such systems are ongoing \cite{bihl2020analytics}.
Historically these functions have been fragmented across military branches \cite{Anthony_Pfaff2023-sr}, but recent initiatives in the US 
have sought to unify military AI and big data platforms \cite{Allen2023-sd}.
One example system is the Joint All-Domain Command and Control program which attempts to automate and speed up the processing of a large variety of sensor, imaging, and intelligence data across branches of the military \cite{Jadc2022-qr}.

A concerning development in this domain recently is Palantir’s 2023 demonstration of its Artificial Intelligence Platform (AIP) for Defense, which uses large language models (LLMs) to recommend military command decisions \cite{Mikhailov2023-su, Michel2023-tm, Reynolds2023-ez}.
Described capabilities include recommendations by the chatbot for deploying specific types of missiles, long range artillery or aerial units in various contexts.
The system’s performance reportedly deteriorated over time, and intervention was needed to restore accuracy and logic in a demo \cite{Mikhailov2023-su}.
Scale AI has also recently announced a partnership with the Center for Strategic and International Studies to create LLMs that can perform wargaming, diplomacy, military command advising, cybersecurity analysis, and even help respond to misinformation campaigns \cite{Scale-AI2024-wf}.
Finally, while no projects have been announced yet, OpenAI recently lifted their ban on using their products for military purposes in early 2024 \cite{Biddle2024-fz}.

A major concern in this domain is that as the complexity of AWS systems and geospatial/intelligence data increase, commanders will rely on annotation, analysis and suggestions from  AI systems whose output cannot easily be double checked \cite{Anthony_Pfaff2023-sr}.
This concern will likely limit proliferation of advanced AI-based command systems in the short term, but as capabilities improve it may be unwisely disregarded.
A recent study has also voiced concern that increased use of AI at the command-level could risk (and may already be) lowering the standards for good tactics and strategy due to how narrow and specialized LLMs inherently are relative to the human brain \cite{Hunter2022-bw}.
Supporting the validity of this concern, a 2024 study of military-related LLMs found that LLMs were prone to recommending pro-escalation tactics with unclear motivation and logic, including escalations that provoked arms races and called for nuclear weapons deployment \cite{Rivera2024-ka}.
In the same vein, concern has recently been expressed about a lack of human oversight in the Israeli ``Lavender'' AI system used for classification of civilians versus militants for military targeting purposes in operations in the Gaza Strip \cite{Abraham2024-wy}. Similar to humans, AI models are not perfect at complex decision making, and more transparency is generally needed in how such systems are used to advise military commanders, particularly regarding what factors are used to set statistical thresholds for lethal decisions.

\section{AWS Proliferation and Threats to Academic Research}
\label{sec:proliferation}

Presently, most forms of AWS are deployed by few countries, and only in limited numbers.
As such, their military significance is relatively low\textemdash AWS capabilities are not core to any nation's national defense at present.
We anticipate this will change over time, possibly within 5-10 years, with AWS taking a more central role in how war is waged.
Further, we expect AWS to proliferate widely, and perform key roles in many armies around the world \cite{Scharre2018-oq}.
If AWS are core to national defense, this means the technologies that power AWS capabilities, machine learning and robotics, are critical defense technologies.
This means that developments in those fields have major military implications, which raises the prospect of national security restrictions on AI research and researchers worldwide.
In this section, we argue why advanced AWS, once developed, will proliferate widely, and why policies that aim to regulate AI research to prevent AWS proliferation will be both futile and deleterious to the field.

Firstly, regarding proliferation, the technologies needed (for example) to allow a tracked UGV such as the TheMIS to maneuver, detect, and fire on enemy troops with a usefully-high success rate already exist as an extension of methods developed for autonomous driving and computer vision applications \cite{Mathiassen2021-lg}.
Likewise, UAVs detecting targets and directing strikes against them without human intervention is already tractable- indeed, multiple loitering munitions advertise some degree of autonomous targeting functionality as described in Section \ref{sec:current_aws}.
While non-trivial integration work is needed to implement such capabilities for any given AWS, the hardware and software engineering talent required is not scarce or restricted, as evidenced by the multitude of AI and ML products being developed by companies worldwide. 

Secondly, ML software proliferates easily and widely and is cheap to develop relative to many military procurement programs \cite{Warren2020-ey, Bengio2023-wn}.
The key breakthroughs enabling advanced AWS now rather than in past decades come from the field of machine learning, where major advances in algorithms have enabled new robotic applications such as warehouse logistics and remote survey and monitoring UAVs and UGVs 
\cite{Schedl2021-nn, Del_Cerro2021-mn}.
In the near future, academic and industry researchers aim to expand that frontier to include human-assistant robots capable of navigating complex indoor environments \cite{Fan2020-xb}, dexterous object manipulation \cite{Andrychowicz2020-tu, Nagabandi2020-vd}, and even surgery \cite{Saeidi2022-ms}, with much effort going into developing such systems.
These are worthy applications, but many of these algorithms are inherently ``dual-use'' in that they can be applied to AWS development by defense contractors \cite{Hoijtink2022-fk, Gomez_de_Agreda2020-bt}.
While it is a good thing that most of the ML software ecosystem remains open source and freely available, this does mean that the core software needed for AWS development is also readily available.

This rapid proliferation of ML talent and open access ML code, while a net positive for humanity, has unfortunately led to recent calls by several top ML researchers for strict regulation, restriction, censorship, or requiring security clearances for many topics in ML  \cite{Bengio2023-wn, Schaake2021-fx, European_Parliament_News2023-bs, Henshall2023-ww, Obrien2023-no}.
Given how vague the criteria for what needs to be regulated are, a paranoid response that censors ML research could have devastating effects on ML progress and global collaboration, and would not limit the proliferation of AWS.
AWS effectiveness and proliferation will likely be limited mostly by the quality and scope of the military datasets available to train them, datasets which have different constraints and goals than civilian datasets \cite{Flynn2020-rk}.
Thus, protecting and securing trained AWS models and relevant datasets is likely a more feasible goal than security compartmentalization of AI research.
We also note that the opportunity to prevent proliferation through locking down the field of ML has largely passed\textemdash
ML expertise and codebases have spread worldwide, and retroactively clamping down on that knowledge would not be sufficient to prevent substantial AWS development from occurring.

One major form these misplaced efforts to restrict knowledge take is immigration and research visa restrictions on civilian scientists.
Indeed, such visa bans have already been put in place for some Chinese researchers by the USA \cite{Hawkins2024-kk}.
Such bans are not just ineffective for the reasons given above, but are pointlessly cruel to the people who are targeted, and can backfire and increase proliferation rather than reduce it.
As a historical example, Qian Xuesen was a Chinese-American rocket scientist who was unjustly expelled from the USA during the early Cold War when rocket science suddenly became a field of critical national security interest, and who as a result became the father of the Chinese rocketry program under coercion after his deportation from the USA \cite{BBC_News2020-if}.

Beyond visa restrictions, attempts to stem the proliferation of potentially dangerous AI capabilities have so far focused on blocking hardware exports to select countries, most notably the US block on GPU exports to China.
However, while economically disruptive \cite{Triolo2023-cy, Allen2023-oh}, blocking access to top-of-the-line GPUs and other compute capacity will not limit the proliferation of AWS.
In China’s example this is partially due to the involvement of domestic academic and tech infrastructure in AWS development, including domestic semiconductor fab technology \cite{Chu_undated-gx, Warren2022-hb, Margarita_Konaev_Ryan_Fedasiuk_Jack_Corrigan_Ellen_Lu_Alex_Stephenson_Helen_Toner_Rebecca_Gelles2023-qv}.

However AWS development, like most robotics currently, does not require high end compute capacity. 
While much attention has accrued to LLMs such as ChatGPT which run on large clusters of expensive and scarce cutting-edge GPUs, most robotics and other edge ML research needs only modest hardware \cite{Chen2019-ko}.
These algorithms and the chips they run on are most relevant to AWS development, because both civilian and military robotics must reckon with limited volume, mass, power, and time budgets for on-robot compute.
The chips suitable for these applications must be both small and low power, running models much smaller than state of the art LLMs, and 
will overlap heavily with those used for common civilian applications.

As a concrete example, Shield AI, who produce the cutting-edge V-Bat AWS platform capable of autonomous swarming behavior, lists the on-platform compute requirements for autonomy as follows: 
``AI Performance: 100 TOPS, GPU: 1024-core Nvidia Ampere architecture GPU + 32 Tensor Cores, CPU: 8-core Arm® Cortex®-A78AE v8.2 64-bit CPU 2MB L2 + 4MB L3'' \cite{noauthor_2023-mt}.
These specifications correspond to Nvidia's Jetson Orin NX edge compute module, which debuted in 2023 on Samsung's 8nm process from 2018, and is representative of the current standard for embedded GPU compute.
The performance of this module falls far below the most recent performance-based export restrictions on GPUs the US government has imposed, with the threshold for export control at 4800 TOPS, over 48 times more powerful than the Jetson Orin NX’s 100 TOPS \cite{Department_of_Commerce2023-yu}.
By comparison, Chinese semiconductor manufacturer SMIC advertises a 7nm-class process, and multiple Chinese silicon design companies have GPU accelerator chips in production or announced, making the production of a Chinese domestic chip equivalent to the Jetson Orin NX likely in the near future.

As AWS technology proliferates and becomes increasingly critical to militaries worldwide, we will likely see nations make use of academic and corporate research talent to aid AWS development.
The example of China’s fusion of military and civilian infrastructure and institutions is unfortunately likely to become a common model for developing states interested in AWS weapons systems \cite{Seraphin2023-xu}.
Maintaining separate well-funded military and civilian research spheres to preserve responsible management of dual-use AI technology is morally admirable, but a significant luxury. 
Several major high ranking Chinese universities have already been blacklisted by the US for contributing to AI weapons development \cite{Margarita_Konaev_Ryan_Fedasiuk_Jack_Corrigan_Ellen_Lu_Alex_Stephenson_Helen_Toner_Rebecca_Gelles2023-qv}, but it is likely universities will be co-opted into AWS development in many other countries over the next decade.
At minimum, it is likely that government funding for academic AI research will increasingly concentrate on dual-use topics that can be applied to AWS development.

Similarly, it is likely that companies with significant civilian AI expertise and infrastructure will see pressure and incentives to contribute to defense contracts and develop technology that is of dual-use interest to the military \cite{Reynolds2023-ez, Biddle2024-fz}.
If the threat posed by AI and AWS continues to be equated in rhetoric to that of nuclear weapons, countries may start restricting ML researchers similar to how nuclear scientists have had their travel restricted and their research on some topics blocked from publication, despite the ineffectiveness of such policies \cite{Daniels2019-ib}.

\section{Policy Recommendations}
\label{sec:policy}

\subsection{Historical AWS Policy Efforts}

To begin this section, we will survey a few major events in AWS regulation and policy over the past two decades. 
Awareness and concern about AWS began to increase after 2001, when the US Congress ordered the Pentagon to make one-third of their ``deep strike'' aircraft fleet unmanned by 2010, and one-third of their ground combat vehicles unmanned by 2015 \cite{Congress2001-pz}.
These timelines proved overly aggressive, but sparked a number of efforts to regulate AWS development\textemdash
In 2010, the International Committee for Robot Arms Control called for regulations on remote-controlled and autonomous weapons \cite{Markoff2010-uy}.
Then, in 2013, the Campaign to Stop Killer Robots started as a coalition of NGOs attempting to push for global bans on lethal AWS \cite{noauthor_undated-xh}.
In 2014, the phrase ``meaningful human control'' (MHC) was coined by Article 36, which has since become a core principle of AWS ethics and regulation \cite{Article2014-vf}. 

Calls for regulation have not been limited to NGOs\textemdash 
A number of national and international efforts to regulate AWS have been seen in recent years.
As of 2020, at least 30 countries have called for a ban on lethal AWS, with many others expressing some degree of concern about AWS \cite{Stauffer2020}.
In the USA, the Block Nuclear Launch by Autonomous Artificial Intelligence Act was introduced in 2023 \cite{Markey2023-zy}.
In 2023 60 countries, including China and the USA, issued a call to action in the Hague towards more responsible use of AI for military applications \cite{Sterling2023-ff}.
Also in 2023, the UN passed a resolution calling for strict regulation on AWS, and for humans to stay in control of lethal decision-making in war \cite{noauthor_undated-sj}.
Lastly, while not directly targeting AWS, the EU has moved to regulate AI risks, first in 2021 and more recently in 2024, with the EU AI Act passed as the first major regulation on artificial intelligence, though its implications for AWS are currently unclear \cite{European_Parliament_News2023-bs}. 

\subsection{International Policy Recommendations}


Here, we will we present several policy recommendations that can help to reduce the negative impacts of AWS, and exhort both ML and policy researchers to work to address these issues and to advocate for their implementation where possible.
Although evidence suggests that the widespread use of AWS is inevitable, the manner in which this comes to pass will play a large role in determining the risks they pose to global stability and to academic freedoms. 

The first policy recommendation we make is that nations can mitigate the negative effects of AWS on conflict risk by maintaining a ban on human-independent use of AWS.
That is, AWS should not be deployed into combat without a meaningful human battlefield presence alongside them, whether using human-AWS teaming processes or as part of separate operations in the same battlespace.
While the widespread deployment of AWS is likely inevitable, there is still time to influence where and how it is deployed, and human-independent deployment is not technically trivial.

There are two avenues to preventing human-independent use: AWS manufacturers should not develop systems designed to operate in an environment free of friendly humans, and policymakers should not field independent combat formations composed entirely of AWS. 
On the technical front, requiring strong human-in-the-loop measures goes a long way, as does limiting the use of generalist AI models in AWS.
The AI research community should be wary of and resist attempts to use large pretrained models in AWS development\textemdash 
These models are likely more useful for building generalist AWS capable of human-independent operations compared to more specialized AWS, which can be developed using smaller and more specialized ML models.
Further, AWS using specialist models will likely prove easier to develop to maturity and more easily fit into existing military command systems and doctrine, so this approach is also less risky from a military procurement perspective.




Our second policy recommendation is that policy experts develop consensus on levels of functional autonomy in AWS, and what constitutes a sufficiently autonomous AWS to require regulation \cite{Unoda2023-bu, Qiao-Franco2023-xd}.
Clear capability and autonomy levels will help to avoid tensions and arms-race dynamics between competing nations as AWS become central to defense planning.
Our ideal would be a full international ban on AWS that make a lethal decision without direct human involvement at the time of the weapon firing, a line that has already been crossed in combat in the Ukraine War \cite{Hambling2023-ps} and by loitering munitions in multiple conflicts \cite{Bode2023-yf}.
Short of this ideal, policymakers need to lead military R\&D by deciding where to draw the line before systems that cross it are developed and deployed in combat, which makes rolling back to lower autonomy levels much harder.

The third policy suggestion we make is for urgent improvements to transparency and oversight regarding planned and deployed AWS capabilities.
In particular, we need clarity on what military tasks are acceptable to automate, and with what level of human supervision, both on paper and in the field.
For example, there is (hopefully) wide consensus that nuclear arsenals should not be turned over to fully autonomous systems exempt from human oversight \cite{Tamburrini2023-ig, Matthew_Robert2021-sw}.
Unfortunately, given recent enthusiasm for autonomous weapons as well as recognition that contexts like signal jamming and underwater warfare inherently benefit from human-out-of-the-loop autonomy, anything less dangerous than autonomous nuclear weapons seems to be under consideration at the moment \cite{Greenwalt2023-io, Anthony_Pfaff2023-sr, ORourke2023-fv}.
All branches of the US military are already investing in command-augmenting or command-replacing AI systems \cite{Jadc2022-qr,Mikhailov2023-su}, but the aims and scope of these projects are vague at best when it comes to the degree of human removal from analysis and targeting.
We support measures to mitigate the worst of these risks, and hope legislation can be pushed to create oversight processes for autonomy in AWS for conventional munitions.

In addition, we note the value of transparency in AWS deployment as battlefields become more and more dominated by machines. 
In such a scenario, there will by default be fewer humans present to report on war time events.
Past AWS-led military excursions have been under-reported to the public and increasingly cross lines that prior political and military statements said would never be crossed \cite{Hernandez2021-aj, Bergman2021-we, Hambling2023-ps, Abraham2024-wy}.
As AWS come to require less human involvement, covert AWS campaigns and secret AWS programs may be known to fewer military and intelligence personnel as well.
This situation may lead to cover-ups of war crimes, as well as insufficient civilian assessment of new systems, leading decision makers to overestimate their capabilities \cite{Anthony_Pfaff2023-sr}.
Providing access to independent watchdogs and journalists should be mandatory for advanced AWS use.
In addition, a robust reporting process which maximizes public transparency should be implemented to assess the outcomes and effectiveness of these AI-based weapons, and to prevent the conduct of war from being further hidden from the public interest.


\subsection{Academic Policy and Action Recommendations}

Lastly, we present several actions for academics and academic institutions to take to address the issues we describe in this work, beginning with an overview of recent actions researchers have taken on AWS regulation.
Most recent academic initiatives have centered on open letters or petitions seeking to regulate or ban AWS \cite{noauthor_2016-bs, Amoroso2020-yt}.
For example, in 2017 a Clearpath Robotics co-founder published an open letter calling for the UN to ban AWS, with 116 founders of robotics and AI companies from 26 countries signing on \cite{Bogdon_undated-oi}.
Similarly, in 2018 AI researchers from 30 countries wrote an open letter protesting a new department at KAIST that merged academic and military research \cite{noauthor_undated-ti}.
In 2018, Google employees' protest led to the company canceling a Pentagon contract that could have led to more accurate drone strike technology \cite{Crofts_Penny2020-di}.
Another Google employee protest has started recently, opposing their Project Nimbus \cite{Perrigo2024-sb}. 
Furthermore, we note that global public opinion is favorable for AWS regulation currently.
A 2020-2021 public opinion survey found that approximately 61 percent of adults across 28 countries oppose the use of fully autonomous weapons, with India being the only country surveyed with majority support for AWS \cite{Ipsos2021-qt}.

Looking forward, AI research may increasingly result in dual-use hardware and software useful for either weapons systems or civilian applications, and expanding military funding for AWS risks more recruitment of AI/ML researchers and institutions onto AWS development projects. 
An immediate way to address this issue is to treat military funding similar to how industry funding is treated by many universities. 
Currently, many universities require substantial ethics training in and provide administrative oversight of industry funding sources.
These measures seek to prevent bias in research or negative effects on the public's perceptions of the role of the university in society \cite{Bickford2004-gb, Robertson2011-qy, Fabbri2018-oz, Larrick2022-qv, Bero2022-sj}.
Military funding is not currently handled with the same degree of oversight and awareness \cite{DAgostino2024-pw}. 
This measure is simple to implement, and can prevent the most ethically questionable AWS research projects and defend academic independence.

We emphasize that this is not a call for universities to swear off military funding altogether.
Defense-funded academic research that benefits military applications diffusely or is disconnected from combat technology is neither new nor a major threat to academic freedom. 
However, the rise of AWS will likely make academic research central to new AI projects for defense and national security that directly lead to more deadly new weapon systems on short timescales.
Given this trend, military funding should be handled with the same awareness and caution that industry funding is currently handled with. 
Clear institutional rules around which military projects are acceptable and which endanger academic independence are needed \cite{DAgostino2024-pw}.

To conclude, unconstrained AWS development poses a range of risks to academics and nations worldwide, many of which have not been properly examined.
It is urgent that we begin to reckon with these threats to our field and society. 

\section*{Acknowledgements}
We are grateful to Samantha Hubner for reviewing and providing feedback on drafts of this work, and for helpful discussion.

\section*{Impact Statement}

This position paper is intimately concerned with the ethical and societal impacts of machine learning.
Thus, in some sense the paper itself is an impact statement about global militaries' recent embrace of ML/AI-based autonomous weapons (AWS), as well as the ML community's growing advertent and inadvertent contributions to AWS advancements.
Regarding the impacts of this paper itself, we hope that they will be both significant and positive, spurring awareness, constructive discussion, and ultimately action from the global ML community and beyond to mitigate the negative effects of AWS.
That said, we do not expect one paper alone to do this. 
It is our hope that this work will encourage further analysis and debate on the impacts of AWS by ML and policy researchers worldwide as well as the public, ultimately leading to action by policymakers in many nations.
We do not anticipate likely negative impacts from this work (such as encouraging ML researchers to become more, rather than less, active in advancing AWS technology), as AWS are already being recklessly developed and deployed.
Furthermore in our opinion, ignoring or limiting awareness of active AWS research carries much greater risks to society than open and honest discussion, and as such it is appropriate and necessary for the field to discuss these issues in a public way.

Last, some may argue that AWS will become essential for national security, and thus that it is unethical to discourage nations from developing AWS because they are needed to maintain military deterrence. 
More broadly, one could ask under what extremes may highly-autonomous AWS development and production be viewed as a necessary evil, and discouraging it therefore unethical?
As a hypothetical example, an aggressive world power may refuse to limit fully-autonomous, human-free AWS development, and over years or decades their technology could progress to the point where it is impossible to counter using conventional forces, human-controlled AWS, or nuclear deterrence.
In this particular situation we concede that fully-autonomous AWS development with a deterrence mentality, mirroring modern nuclear deterrence and tenuous geopolitical equilibriums, may pragmatically become a necessity.
However, as outlined in this paper, while AWS battlefield deployments have started in preliminary ways, AWS are currently not the core of any nation's military, and global public and institutional opinions are largely in favor of substantial regulation of some form.
As such, this hypothetical is far from guaranteed to come to pass.
However, we do not believe that the contributions of this paper are contingent on taking a position regarding the above hypothetical.
Our work concerns the impacts of AWS on conflict risk and academic freedom, which will occur regardless of whether AWS are under strict human control or are highly autonomous, so long as they are critical to warfare and replace the human battlefield element. 
The goal of this work is to open discussion on the impacts of AWS among AI researchers and ground it in prior work, and we expect that the ethics of meaningful human control will be a major topic of future debate in the community, but our contribution does not rely on taking an ethical position on the matter.

\bibliography{references}
\bibliographystyle{icml2024}

\newpage
\appendix
\onecolumn

\section{Global Regions with Alleged AWS Deployment Discussed in This Work}

\begin{figure}[htp]
    \centering
    \includegraphics[width=15cm]{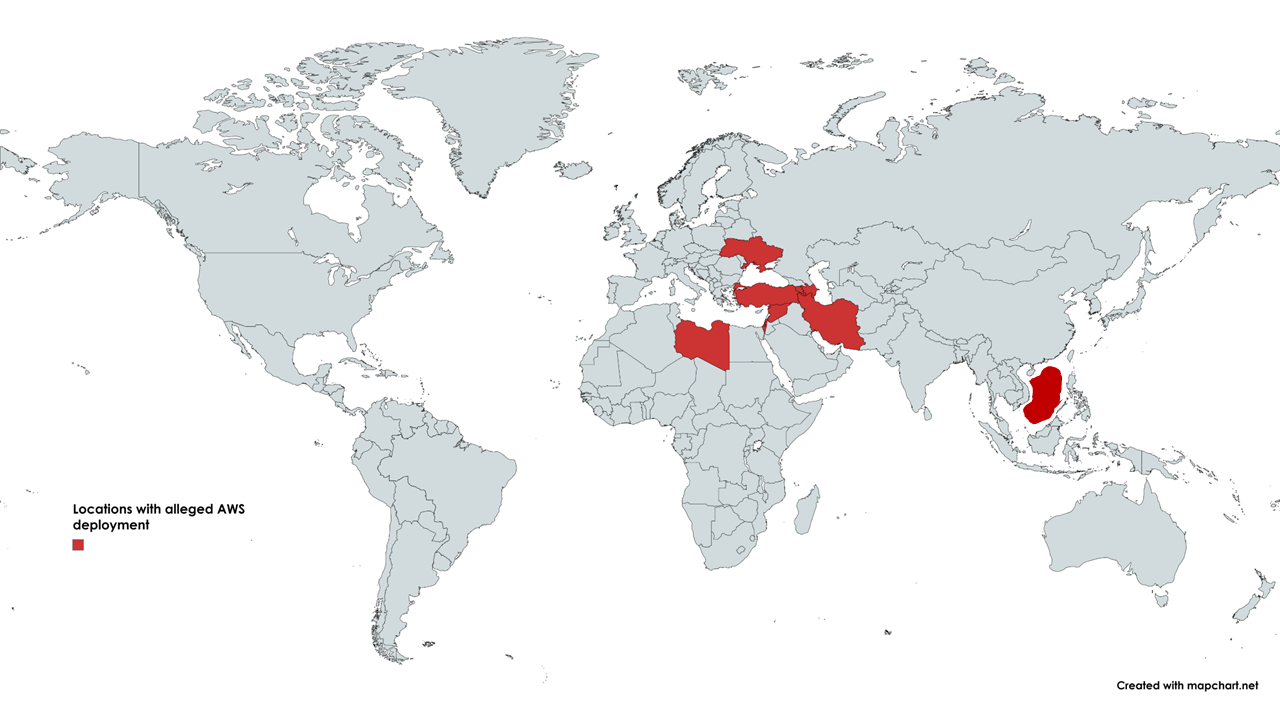}
    \caption{Locations of alleged deployment of fully or near fully autonomous AI-powered weapon systems that were discussed in this work. The Turkey, Azerbaijan and Armenia deployments are further discussed by \cite{De_Vynck2021-qa}.}
    \label{fig:galaxy}
\end{figure}



\end{document}